\documentclass[english,12pt]{revtex4}

\usepackage{graphicx}

\begin{document}

\author{Novoselov A.A.}\email{Alexey.Novoselov@cern.ch}
\affiliation{Institute for High Energy Physics, Protvino, Russia}
\affiliation{Moscow Institute of Physics and Technology, Dolgoprudny, Russia}
\title{A study of charm hadron production in $e^+e^-$ annihilation}

\begin{abstract}
The processes of $D^{(*)}$-mesons and $\Lambda_c$-baryons production in $e^+e^-$-annihilation
at $10.58$ GeV and $91.18$ GeV energies are concerned.
At the $10.58$ GeV energy the production of charmed particles via the $B$-mesons
decays is also concerned.
Scaling violation of the fragmentation functions is calculated at the
NLL-accuracy.
Nonperturbative fragmentation functions are retrieved from the experimental
data of $B$-factories and are approximated by simple analytic expressions.
It is proved, that the difference between nonperturbative fragmentation functions
of mesons and baryons can be easily explained by quark counting.
\end{abstract}

\maketitle

\section{Introduction}

The production of heavy quarks (charm and bottom) in the high-energy collisions and their
subsequent hadronization are important processes in particle physics. Apart from being
of interest per se,
these processes are significant for some practical purposes. For
instance, a light Standard Model Higgs boson is expected to decay preferably into
heavy quark-antiquark pairs, which then fragment into heavy hadrons. The direct
production of heavy quarks would, thereby, be the main background process.
Understanding of hadronization processes is also important for the precise determination
of the mass of the decaying particle.

Finite heavy-quark mass $m_Q$, providing a natural infrared cutoff, allows to carry out
the calculations to a large extent using perturbative QCD. Nonetheless, differential
momentum distributions of the heavy hadrons produced in high-energy collisions are sensitive
to the large logarithms $\ln{s/{m_Q^2}}$, where $s$ is the center-of-mass energy squared.
Since for modern colliders $\sqrt{s}\gg m_Q$, these logarithms
threaten the convergence of perturbation
expansions. Fortunately, it is possible to show that up to the power corrections in
$m_Q^2/s$ the cross section factorizes into mass-independent hard partonic cross
sections convoluted with so-called fragmentation functions describing the probability for
the parton to fragment into a particular heavy hadron
\cite{Mele:1990cw,Collins:1998rz}.

Let us consider the inclusive production of a heavy hadron $H$
via the decay of a vector boson $V=\gamma^*,Z^0$ produced in the $e^+e^-$ annihilation:
\begin{equation}\label{reaction}
   e^+ e^-\to V(q)\to H(p_H)+X .
\end{equation}

It is convenient to introduce the scaling variable $x$ for expressing
the heavy hadron energy in the center-of-mass frame,
\begin{equation}\label{xdef}
   x = \frac{2p_H\cdot q}{q^2}
   \stackrel{\rm C.M.}{=} \frac{2E_H}{\sqrt s}.
\end{equation}
Experimental data are often presented in terms of the scaled momentum
$x_p=\left|\vec{p}_H\right|/\sqrt{s/4-m_H^2}$.
But due to large energy of interaction both scaling variables
are almost indistinguishable.

As stated above, at leading power in $m_Q^2/s$ the differential production
cross section can to all orders of perturbation theory be factorized as
\begin{equation}\label{fact}
   \frac{d\sigma_H}{dx} = \sum_a
   \frac{d\hat\sigma_a}{dx}(x,\sqrt{s}) \otimes
   D_{a/H}(x,m_Q,\mu) ,
\end{equation}
where $d\hat\sigma_a/dx$ is the cross section for producing a massless parton $a$
with the scaled energy $x$ after subtracting the collinear singularity in the
$\overline{\rm MS}$ factorization scheme,
and fragmentation function $D_{a/H}$ gives the probability for a parton $a$
to fragment into
a heavy hadron $H$ carrying a fraction $x$ of the parton's momentum.
$D_{a/H}$ also depend on the factorization scheme,
but the convolution of the two term is not, so that the physical cross section
is prescription independent \cite{Mele:1990cw}.

The factorization formula (\ref{fact}) separates the dependence on the heavy-quark
mass $m_Q$ from the dependence on the center-of-mass energy which is contained
in the partonic cross sections.
These hard cross sections can be calculated in the massless approximation.
All dependence on the resulting hadron
resides in the fragmentation functions
which are process-independent non-perturbative quantities.
Like the parton distribution functions (PDF's) fragmentation functions must be measured at some scale and
their values at any other scale can be obtained by solving the DGLAP
evolution equations \cite{Dokshitzer:1977sg,Gribov:1972ri,Altarelli:1977zs}.

The fragmentation functions incorporate long-distance, non-perturbative
physics of the hadronization process in which observed hadrons
are formed
from the partons. For proper understanding of hadronic uncertainties
one needs to separate short $p \sim q$ and long $p \sim \Lambda_{\rm QCD}$
distance effects. The most popular approach is to factorize the fragmentation
function into perturbative and non-perturbative components:
$D_{a/H}=D_{a/H}^{\rm pert}\otimes D_{Q/H}^{np}$
\cite{Colangelo:1992kh,Cacciari:1996wr,Nason:1999zj}.
The first component is identified with the so-called perturbative fragmentation function,
$D_{a/H}^{\rm pert}(x,m_Q,\mu)=D_{a/Q}(x,m_Q,\mu)$ while for the non-perturbative component
a model such as Kartvelishvili et al. \cite{Kartvelishvili:1977pi,Kartvelishvili:1978jh} or
Peterson et al. \cite{Peterson:1982ak} is adopted.

Perturbative fragmentation functions
$D_{a/Q}$ describe the probabilities for partons $a$ to fragment into an on-shell heavy quark $Q$.
They are relevant for the discussion of inclusive heavy-quark production, where one sums over
all possible hadron states $H$ containing heavy quark $Q$.
Quark-hadron duality suggests that
\begin{equation}
   \sum_H D_{a/H}(x,m_Q,\mu) = D_{a/Q}(x,m_Q,\mu)
   + {\cal O}\bigg( \frac{\Lambda_{\rm QCD}}{m_Q(1-x)} \bigg) .
\end{equation}

However, such a relation can be expected to hold only if $x$ is not too close to 1,
so that the scale $m_Q(1-x)$
is in the short-distance regime \cite{Neubert:2007je}.
It is important to mention, that such an ansatz spoils the proper factorization
of short- and long-distance contributions. For $x\to 1$ perturbative fragmentation
function itself contains long-distance contributions from logarithms of momentum
scales of order $m_Q(1-x)\sim\Lambda_{\rm QCD}$.
Such logarithms are responsible for soft gluon emission and are not controllable in
the perturbation theory.
The previous attempts to resum the $\ln^n(1-x)$ terms in the fragmentation functions have
thus led to unphysical negative values in the $x\to 1$ region \cite{Cacciari:2005uk}.
As a matter of fact, the Landau singularity of the perturbative coupling at small
transverse momenta leads to a branch-point in the resummed expression
and produces the negative behavior at $1-x \sim \Lambda_{\rm QCD} / m_Q$.
Soft-gluon resummation therefore suggests that the non-perturbative
phenomena become dominant when $x > x_{br} \approx 1 - \Lambda_{\rm QCD} / m_Q$.
The invariant mass of heavy quark and soft gluons
emitted can be estimated as
$m_Q(1 + (1-x_{br})/{x_{br}}) \approx m_Q / (1 - \Lambda_{\rm QCD} / m_Q) \approx m_Q + \Lambda_{\rm QCD}$.
It means that soft gluons revealing themselves at $\Lambda_{\rm QCD}$ scale can play
appreciable role in the hadronization process.

To incorporate non-perturbative hadronization effects into the heavy-quark
fragmentation functions is the main objective of this work.
Thus, let us not to resum to all orders of
perturbation theory the long-distance terms in the perturbative contribution.
The motivation against resummation was adduced above.
The non-perturbative fragmentation function will be numerically
retrieved from experimental data. Being retrieved from the $B$-factories data
on the $D^*$ production at the $\Upsilon(4S)$ energy, this function
will allow to describe the ALEPH data at $Z$-boson peak with the reasonable precision. Apart from
testing evolution of the $D^*$ fragmentation the difference between meson and baryon fragmentation
will be studied. It will be shown that in the $x \to 1$ region this difference
is in agreement with the premises of Kartvelishvili et al. model.


\section{Theoretical preliminaries}

\subsection{Perturbative fragmentation function and QCD evolution}

With the use of factorization relation for fragmentation function
and the assumption that $D_{a/H}^{\rm pert}(x,m_Q,\mu)=D_{a/Q}(x,m_Q,\mu)$
eq. (\ref{fact}) can be rewritten as follows:
\begin{eqnarray}
\label{fact2}
   \frac{d\sigma_H}{dx}(x,\sqrt{s},m_Q) &=&  \sum_a
   \frac{d\hat\sigma_a}{dx}(x,\sqrt{s}) \otimes
   D_{a/Q}^{\rm pert}(x,m_Q,\mu)\otimes D_{Q/H}^{\rm np}(x) = \nonumber \\
   &=&\frac{d\sigma_Q}{dx}(x,\sqrt{s},m_Q)\otimes D_{Q/H}^{\rm np}(x) ,
\end{eqnarray}
where ${d\sigma_Q}/{dx}$ is the heavy quark differential inclusive cross
section.

The $\overline{\rm MS}$ fragmentation functions  $D_{a/Q}$ obey the DGLAP
evolution equations
\begin{equation}
\label{eq:AP}
  \frac{d D_{a/Q}}{d\ln\mu^2} (x,m_Q,\mu) = \sum_b\int^1_x \frac{dz}{z}
  P_{b  a}\left(\frac{x}{z},\alpha_s (\mu) \right) D_{b/Q}(z,m_Q,\mu) .
\end{equation}
Let us introduce a notation
\begin{equation}
\bar{\alpha}_s(\mu)=\frac{\alpha_s(\mu)}{2 \pi} ,
\end{equation}
where the standard two-loop expression for $\alpha_s(\mu)$ is used.
Then the perturbative expansion for the Altarelli-Parisi splitting functions $P_{b a}$
would have the following form:
\begin{equation}
\label{eq:APfunctions}
    P_{b a}\left(x,\bar{\alpha}_s(\mu)\right) = \bar{\alpha}_s(\mu) P^{(0)}_{b a}(x)
    + \bar{\alpha}_s^2(\mu) P^{(1)}_{b a}(x) + {\cal O}(\bar{\alpha}_s^3) ,
\end{equation}
where the $P_{b a}^{(0)}$ are
\cite{Altarelli:1977zs}
\begin{eqnarray}
P^{(0)}_{QQ}(x)&=&C_{F}\left[\frac{1+x^2}{(1-x)_+}
+\frac{3}{2}\delta(1-x)\right]\;,
\nonumber \\
P^{(0)}_{gg}(x)&=&2 C_{ A}\left[
\frac{x}{(1-x)_+}+\frac{1-x}{x}+x(1-x)+
\left(\frac{11}{12}-\frac{n_f T_{F}}{3 C_{A}}\right)
\delta(1-x)\right]\;,
\nonumber \\
P^{(0)}_{gQ}(x)&=&C_{F}\frac{1+(1-x)^2}{x}\;,
\nonumber \\
P^{(0)}_{Qg}(x)&=& T_{F}\left[x^2+(1-x)^2 \right]\;,
\end{eqnarray}
The NLO splitting functions $P_{ji}^{(1)}$ (needed to achieve NLL accuracy)
have been computed in
\cite{Curci:1980uw,Furmanski:1980cm,Floratos:1981hs,Kalinowski:1980we,Kalinowski:1980ju}
and are too lengthy to be replicated here.

The initial conditions for the $\overline{\rm MS}$  fragmentation functions
were first obtained at the NLO level in \cite{Mele:1990cw}.
They are given by
\begin{eqnarray}
{D}^{\rm ini}_{Q/Q}(x,m_Q,\mu_0)&=&\delta(1-x)+\bar{\alpha}_s(\mu_0)\,
d^{(1)}_Q(x,m_Q,\mu_0)+{\cal O}(\bar{\alpha}_s^2)\;,
\nonumber \\
\label{eq:Dini}
{D}^{\rm ini}_{g/Q}(x,m_Q,\mu_0)&=&\bar{\alpha}_s(\mu_0)\, d^{(1)}_g(x,m_Q,\mu_0)+
{\cal O}(\bar{\alpha}_s^2)\;,
\end{eqnarray}
(other $D_{a/Q}$ are of order $\alpha_s^2$), where
\begin{eqnarray}
d^{(1)}_Q(x,m_Q,\mu_0)&=&C_F\left[
\frac{1+x^2}{1-x}\left(\ln\frac{\mu_0^2}{m_Q^2}-2\ln(1-x)-1\right)\right]_+
\nonumber  \,,
\\
d^{(1)}_g(x,m_Q,\mu_0)&=& T_F \left[ x^2+(1-x)^2 \right] \ln \frac{\mu_0^2}{m_Q^2}\;.
\end{eqnarray}

Although the sum in expression (\ref{fact2}) runs over all types of partons,
$D_{g/Q}$ is $\alpha_s$-suppressed with respect to $D_{Q/Q}$ while other $D_{a/Q}$
being $\alpha_s^2$-suppressed. So in the following let us keep
only direct component $D_{Q/Q}$ since it is quite sufficient for the purposes
of current work. Thus, expanding the convolution, for heavy quark spectrum one obtains
\begin{eqnarray}
\label{factQ}
   \frac{d\sigma_Q}{dx}(x,\sqrt{s},m_Q,\mu) &=&
   \int^1_x \frac{dz}{z}
   \frac{d\hat\sigma_Q}{dz}(x/z,\sqrt{s})
   D_{Q/Q}(z,m_Q,\mu) ,
\end{eqnarray}
where NLO expression for the partonic cross section
from \cite{Nason:1994xx} should be used:
\begin{eqnarray}
\frac{d\hat\sigma_Q}{dx}(x,\sqrt{s}) &=& \delta(1-x) + \bar{\alpha}_s(\mu) a^{(1)}_{Q}(x,\sqrt{s}),
\nonumber
\\
a^{(1)}_{Q}(x,\sqrt{s}) &=&
C_F \left[ 1 +
\ln { s \over m^2 } { \left( { {1+x^2} \over {{(1-x)}_+} }  + { {3} \over {2} }  \delta(1-x) \right) } +
\right.
\nonumber \\
&+& {1\over 2} { {x^2 - 6 x - 2} \over {(1-x)_+} } - {\left({{\ln(1-x)}\over{1-x}}\right)}_+ (1+x)^2 +
\nonumber \\
\label{coeff-func}
&+&
\left.
2 {{1+x^2}\over{1-x}} \ln x
+ \left( {2 \over 3} \pi^2 - {5 \over 2} \right) \delta(1-x)
\right].
\end{eqnarray}

The procedure outlined above guarantees that all leading
and next-to-leading logarithmic terms of quasi-collinear origin
(terms of the form
$(\bar{\alpha}_s \log(q^2/m_Q^2))^n$ and $\bar{\alpha}_s (\bar{\alpha}_s \log(q^2/m_Q^2))^n$ respectively)
are correctly resummed in the cross section \cite{Mele:1990cw}.

For the subsequent analysis
it is convenient to turn to the Mellin moments of the quantities involved.
The Mellin transformation $f(N)$ of function $f(x)$ is defined as
\begin{equation}
   f(N) \equiv \int_0^1 dx \, x^{N-1} f(x)\;.
\end{equation}

In Mellin space the evolution equations (\ref{eq:AP}) take the simple form
\begin{eqnarray}
\label{dlgap-N}
\frac{dD_{a/Q}}{d\ln\mu^2}(N,m_Q,\mu) = \sum_b
P_{ba}(N,\alpha_s (\mu)) D_{b/Q}(N,m_Q,\mu)\;.
\end{eqnarray}

This equation at NLO level
was solved analytically in \cite{Mele:1990cw} and for the direct component $D_{Q/Q}$ one has:
\begin{eqnarray}
D_{Q/Q}(N,m_Q,\mu) &=& E(N,\mu,\mu_0)  D^{\rm ini}_{Q/Q}(N,m_Q,\mu_{0}) , \nonumber \\
E(N,\mu,\mu_0) &=&
\exp\Bigg\{
\ln\frac{\alpha_s(\mu_{0})}{\alpha_s(\mu)}\; \frac{P_{QQ}^{(0)}(N)}{2\pi b_0} +
\nonumber \\
&+&
\frac{ \alpha_s(\mu_{0}) - \alpha_s(\mu) }{4\pi^2 b_0}
\left[ P_{QQ}^{(1)}(N) - \frac{2\pi b_1}{b_0} P_{QQ}^{(0)}(N) \right]
\Bigg\} .
\label{evosol}
\end{eqnarray}

Defining
\begin{eqnarray}
\sigma_c(N,\sqrt{s}) \equiv \int_0^1 dx \, x^{N-1}
\frac{d\sigma_c}{dx}(x,\sqrt{s})\;,
\end{eqnarray}
one has the following expression for the NLO distribution:
\begin{eqnarray}
\sigma_c(N,\sqrt{s}) =
{\hat\sigma_Q}(N,\sqrt{s})
\, E(N,\mu,\mu_0)
\, D_{Q/Q}^{\rm ini}(N,\mu_0,m_Q).
\label{sigma-pff-N}
\end{eqnarray}

Both $a^{(1)}_Q$ and $d^{(1)}_Q$ contain terms proportional to the
$\alpha_s / (1-x)_+$ and $\alpha_s \left[ \ln(1-x) / (1-x) \right]_+$,
associated to the emission of a soft gluon. These terms give rise to a
large-$N$ growth of the corresponding Mellin transforms
\begin{eqnarray}
\label{eq:a_Q^1}
{a}_Q^{(1)}(N,\sqrt{s},\mu) &=& 
C_F \left[  \ln^2N
+ \left(\frac{3}{2}+2\gamma_E - 2\ln\frac{s}{\mu^2}\right) \ln N  + \alpha_Q
+ {\cal O}(1/N) \right] \!,
\nonumber \\
\label{eq:d_Q^1}
d_Q^{(1)}(N,\mu_0,m_Q) &=& 
 C_F \!\left[ - 2\ln^2N +
2\!\left( \ln \frac{m^2}{\mu_0^2} - 2\gamma_E + 1\! \right)\!\ln N +
 \delta_Q  + {\cal O}(1/N) \right]\! .
\end{eqnarray}

Leading $\alpha_s^n \ln^{n+1}N$ and next to leading $\alpha_s^n \ln^{n}N$ logarithmic
contributions were resummed to all orders of perturbation theory in \cite{Cacciari:2001cw}.
Opposed to fixed order calculation, which leads to finite and positive
fragmentation function at almost all values of $x$ except the $x \to 1$ region
(where the $\delta$-function from the initial condition becomes apparent),
NLL resummed result exhibit pathological
negative behavior when $x$ approaches $1$.
The reason is that the Landau singularity of the perturbative QCD coupling at small
transverse momenta leads to branch-points in the resummed expression for the initial condition
and the coefficient function.

In the initial condition the singularities start at the branch-point
\begin{equation}
N^L_{{\rm ini}}=\exp \left( \frac{1}{2\,b_0
\alpha_s(\mu_0)}\right) \simeq \frac{\mu_0}{\Lambda_{\rm QCD}}
\label{eq:Nini} \;,
\end{equation}
while for the coefficient functions the branch-point is
\begin{equation}
N^L_q=\exp \left(\frac{1}{b_0 \alpha_s(\mu)}\right) \simeq
\frac{\mu^2}{\Lambda_{\rm QCD}^2}\;.
\label{eq:NL}
\end{equation}

The previous attempts to restore the physical behavior of fragmentation functions
consisted in introducing a tower of power corrections to $N$
represented by the replacement
\begin{equation}
 N  \to N\,\frac{1+f/N^L_{\rm ini}}{1+f\,N/N^L_{\rm ini}}\;,
\label{eq:N_tilde_ini}
\end{equation}
in the initial condition and
\begin{equation}
 N  \to N\,\frac{1+f/N^L_q}{1+f\,N/N^L_q}\;,
\label{eq:N_tilde}
\end{equation}
in the coefficient function. It is easy to see that $f$ being more or equal
to $1$ unphysical region is unreachable. But it is important to keep in mind
that there is no rigorous justification for the replacements (\ref{eq:N_tilde_ini}), (\ref{eq:N_tilde}).

\subsection{Non-perturbative fragmentation function}

The most popular parameterizations for the non-perturbative fragmentation functions
are Peterson et al. \cite{Peterson:1982ak} and Kartvelishvili et al. (KLP)
\cite{Kartvelishvili:1977pi,Kartvelishvili:1978jh}.
The former has a form of the heavy quark propagator and does not depend on the hadron produced.
Thus let us concentrate on the latter one.
It is based on the Gribov-Lipatov ``reciprocity relation" between $D_{Q/H}^{np}$ and the distribution function
of quark $Q$ in hadron $H$ \cite{Gribov:1972ri}:
\begin{eqnarray}
D_{Q/H}^{\rm np}(z)&\stackrel{z \to 1}{=}&f_H^Q(z),
\label{eq:reciprocity-relation}
\end{eqnarray}
where $z=p_H / p_c$ is the hadron momentum fraction with the respect to the heavy quark momentum.
The expression for $f_{D^*}^c$ was found out in \cite{Chliapnikov:1977fc} on the basis of Kuti-Weisskopf
model. The parametrization obtained is significantly related to the Regge trajectory
parameters of the $Q \bar Q$-system and has the following form:
\begin{eqnarray}
f_{D^*}^c(x)&=& \frac{\Gamma(2+\gamma_M-\alpha_Q-\alpha_q)}{\Gamma(1-\alpha_Q)\Gamma(1+\gamma_M-\alpha_q)}  x^{-\alpha_Q}(1-x)^{\gamma_M-\alpha_q},
\label{eq:npFF-D}
\end{eqnarray}
where $\alpha_q=1/2$ is the intercept of the light quarks trajectory $\rho, \omega, f, A_2$,
$\alpha_Q$ is the
intercept of the leading trajectory for the $Q \bar Q$-system and $\gamma_H$ is a parameter
determining the behavior of the distribution function for $x \to 1$.
Its value originates from the $q^{(-2)}$ diminution of the form-factor as
it is known that $q^{(-2 k)}$ diminution of form-factor leads to the
$(1-x)^2k-1$ behavior of the distribution function at large $x$.
Assuming the universality
of the sea quarks distribution in all mesons one gets $\gamma_M=3/2$.
In much the same way for
$\Lambda_C$-baryons the following expression was found out:
\begin{eqnarray}
f_{\Lambda_c}^c(x) &=& \frac{\Gamma(3+\gamma_B-\alpha_Q-2 \alpha_q)}{\Gamma(1-\alpha_Q)\Gamma(1+\gamma_B-2 \alpha_q)}
 x^{-\alpha_c}(1-x)^{1 + \gamma_B - 2 \alpha_q},
\label{eq:npFF-B}
\end{eqnarray}
where $\gamma_B=3$ and the factor $2$ is related to the number of valent light quarks in baryon.

Gribov-Lipatov ``reciprocity'' relation with known distribution functions
(\ref{eq:npFF-D}), (\ref{eq:npFF-B}) determine the behaviour of the fragmentation functions
in the $x \to 1$ limit. For the small values of $x$ the condition for the fragmentational approach
to be valid ($p \gg m_Q$) is not satisfied. Nonetheless, if at small $x$ values the production
of heavy hadrons mainly depend on their wave functions, then coinciding $x \to 1$ asymptotes
of (\ref{eq:npFF-D}) and (\ref{eq:npFF-B}) prognosticate similar behaviour of the
momentum distributions of mesons and baryons
in this region.

There is still some uncertainty in the value of $\alpha_c$. Theoretical investigations \cite{Gershtein:2006ng}
based on Regge trajectory systematics result in the value for $\alpha_c$ in the range
between $-2.0$ and $-3.5$. It is slightly more than the value of $\alpha_c \approx -3 \div -4$,
obtained in \cite{Khodjamirian:1992pz} with the use of the QCD sum rules.
Another way to determine $\alpha_c$ by the value of the heavy quarkonia wave function in
the center point leads to the value $-3.5 \pm 0.6$ \cite{Kartvelishvili:1985ac}.

\section{Charm hadrons data fits near the $\Upsilon(4S)$}

High quality data on the charmed hadron production is
provided by BELLE, BABAR and CLEO collaborations
\cite{Aubert:2006cp,Seuster:2005tr,Artuso:2004pj}.
One thus has the opportunity to
perform a more accurate fit for the non-perturbative initial conditions.
Furthermore, it gives us the possibility to test the evolution of the fragmentation function
from the center-of-mass energies of $10.6$ to $91.2~{\rm GeV}$, using charm data
from the LEP experiments \cite{Barate:1999bg,Ackerstaff:1997ki}.

For the perturbative component of fragmentation function we use expression (\ref{sigma-pff-N})
with the  NLO initial condition (\ref{eq:Dini}), the NLO partonic cross section (\ref{coeff-func})
and the NLL evolution (\ref{evosol}) \footnote{Actually an expression analogous to (4.15) in \cite{Mele:1990cw}
is used for calculations.}.
As stated above we do not perform NLL Sudakov resummation
for the initial conditions and coefficient functions retaining these long-distance
contributions for the non-perturbative component.

Several parameters enter the perturbative calculations. First of all these are
the initial $\mu_0$ and the final $\mu$ evolution scales which allow variation by
a factor of order of $2$ around $m_c$ and $\sqrt{s}$ respectively. We set
them to be $\mu_0 = 2 m_c$ and $\mu = \sqrt{s} / 2$ as this values allow to
perform the most successful fiting.
The center-of-mass energy $\sqrt{s}$ is equal to $m_{\Upsilon(4S)}=10.58~{\rm GeV}$.
We shall use
the pole mass for charm quark and fix it to be $m_c=1.6~{\rm GeV}$.
As the $b$-quark mass lies between $\mu_0$ and $\mu$ it creates
a threshold on the different sides of which different number of active flavours
enter the evolution operator. We set $m_b=5.0~{\rm GeV}$.
Experimental value of $\alpha_s(m_Z)=0.119$ points to the
value of $\Lambda_{\rm QCD}$ equal $0.226~{\rm GeV}$.

There are several ways to retrieve the non-perturbative component. DGLAP equations
allow to perform the evolution from a larger scale to a lower one as well as in the
opposite direction. But then
one needs to extract the non-perturbative function from its convolution with the
partonic cross section and the perturbative initial condition. This procedure is
connected with the inverse Mellin transform of the moments of experimental data
divided by the moments of the the partonic cross section and the perturbative initial condition.
Such calculation performed numerically faces problems with the integral's convergence
developing into unphysical negative values of the non-perturbative function in the low $x$
region. It was tested that resulting non-perturbative function obtained by such way
does not permit to
successively reproduce the experimental data by the reverse procedure.

Anther method, proposed in this work, is to represent the function required as a linear combination
of functions which have a simple analytic form of Mellin transform. Further,
expansion coefficients can be determined by fitting to experimental data in
the Mellin space as well as in the $x$-space. The most general choice is
to retrieve the non-perturbative fragmentation function bin-by-bin, choosing the
number of bins $n$ to be not more than the number of experimental points
to avoid overdetermined system for the coefficients. So, let us define
\begin{equation}
\widetilde D^{\rm np}(z) = \sum_{i=1}^n { c_i \, \Theta\left(z-\frac{i-1}{n}\right)\Theta\left(\frac{i}{n}-z \right)}.
\end{equation}
The corresponding Mellin transform is
\begin{equation}
\widetilde D^{\rm np}(N) = \sum_{i=1}^n { c_i \, \int_{\frac{i-1}{n}}^{\frac{i}{n}} {z^{N-1} dz} } =
\sum_{i=1}^n { c_i \, \frac{(i/n)^N-((i-1)/n)^N}{N}}.
\end{equation}

The approximate number of points in the BELLE and BABAR data is 50, in the CLEO data --- 20.
Thus to take 20 bins for the non-perturbative function required seems to be a reasonable choice.

\begin{figure}%
\centering
\parbox{2.9in}{%
\includegraphics[width=3in]{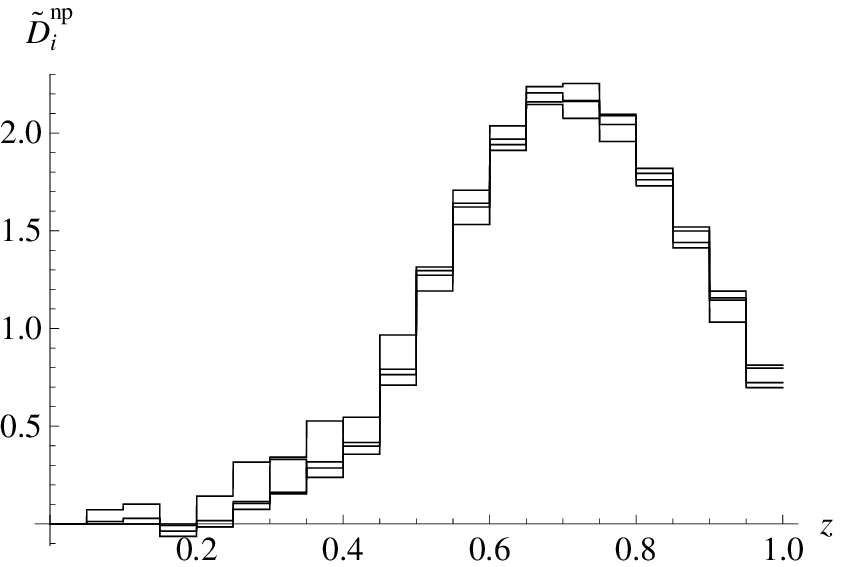}
\caption{\footnotesize Nonperturbative fragmentation functions of the $D^*$-mesons extracted from several datasets. }%
\label{fig:numDstar}}%
\qquad
\begin{minipage}{2.9in}%
\includegraphics[width=3in]{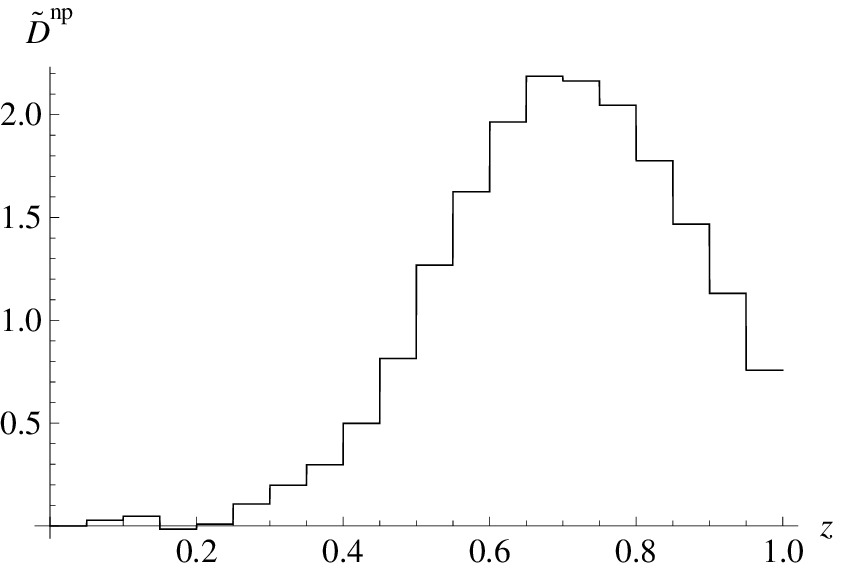}
\caption{\footnotesize Averaged nonperturbative fragmentation function of the $D^*$-mesons.}%
\label{fig:numDstarAv}%
\end{minipage}%
\end{figure}

Fit to the experimental data in the $N$-space can be performed as well as in the
$x$-space. The $c_i$ coefficients obtained by both techniques are in a good agreement.
There are 4 data-sets for $D^*$ production at $\Upsilon(4S)$ energies. Two of
them presented by BELLE and two by CLEO, they regard to the $D^{*+}$ and $D^{*0}$ production.
The non-perturbative fragmentation functions extracted from them are plotted
in Fig. \ref{fig:numDstar}.
As we assume that there is no difference between the $D^{*+}$ and $D^{*0}$ fragmentation
these functions can be averaged to get the $D^{*}$ non-perturbative fragmentation function
plotted in Fig. \ref{fig:numDstarAv}. The weights in this average were selected proportional
to the statistics gathered for each data-set.

\begin{figure}%
\centering
\parbox{2.9in}{%
\includegraphics[width=3in]{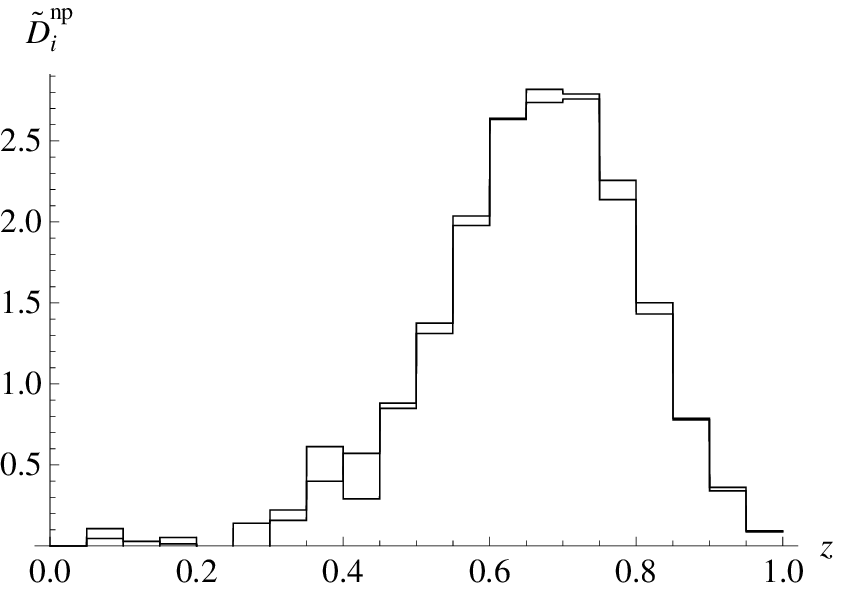}
\caption{\footnotesize Nonperturbative fragmentation functions of the $\Lambda_C$-baryons extracted from several datasets. }%
\label{fig:numLc}}%
\qquad
\begin{minipage}{2.9in}%
\includegraphics[width=3in]{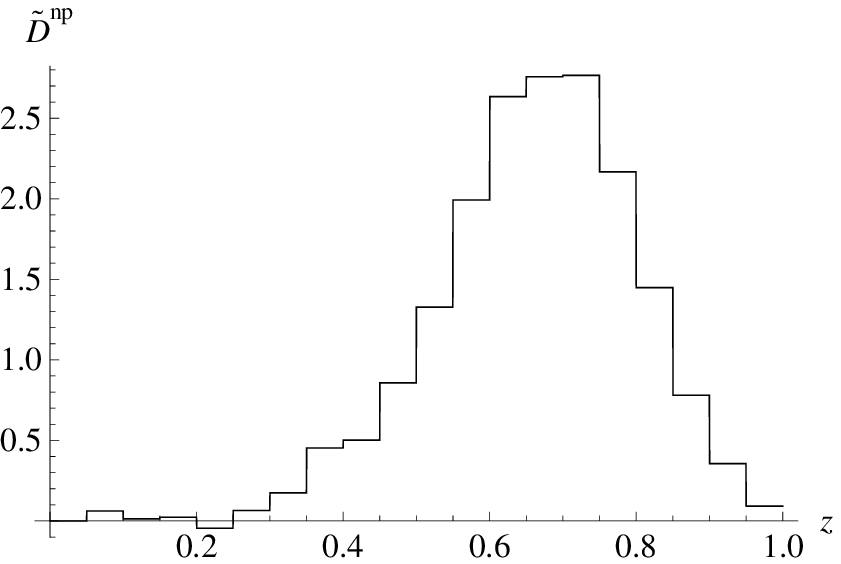}
\caption{\footnotesize Averaged nonperturbative fragmentation function of the $\Lambda_C$-baryons. }%
\label{fig:numLcAv}%
\end{minipage}%
\end{figure}

The same procedure was carried out for the $\Lambda_C$-baryon production at BABAR and BELLE.
The corresponding plots for the non-perturbative fragmentation functions are presented in
Fig. \ref{fig:numLc} and \ref{fig:numLcAv}.

The mesonic and baryonic fragmentation functions obtained in such way do not
reveal significantly different behaviour at $z \lesssim 0.5$ (Fig. \ref{fig:numMBcmp}).
Furthermore, best fits by $c \cdot x^{-\alpha}$ function in the $x < 0.5$ region for both
cases result in close values of $\alpha$.
These values being equal to $-3.7$ and $-3.8$ are
in a pretty good agreement with the previously mentioned predictions for $\alpha_c$.
What concerns the $z \to 1$ behaviour, the difference in it is in agreement
with the Gribov-Lipatov ``reciprocity'' relation.

For further phenomenological analysis it is convenient to find some simple expression
approximating the numeric data obtained:
\begin{eqnarray}
{\widehat D}^{D^*}_c(z) &=& 20.1 z^{3.7}  (1-z) + 2.77  \, {10^3}  z^{13} (1-z)^{7},
\label{eq:npFF-klpM} \\
{\widehat D}^{\Lambda_c}_c(z) &=& 72.9 z^{3.7} (1-z)^5+2.93 \, {10^4} z^{10} (1-z)^5 + 10^3 z^{10} (1-z)^3 .
\label{eq:npFF-klpB}
\end{eqnarray}
Each of them has the same $z \to 0$ and $z \to 1$ asymptotes as the corresponding
KLP function.

\section{Charmed hadron production in annihilation processes}

According to the formula (\ref{fact2}) moments of the momentum spectrum of the particles
produced are equal to the product of moments of the differential partonic cross-section
(\ref{sigma-pff-N}) and moments of the corresponding non-perturbative fragmentation
function. To return to the $x$ variable the inverse Mellin-transform should be performed
by integrating over the vertical line in a complex plane:
\begin{eqnarray}
\label{eq:sigm-x}
\frac{d\sigma_H}{dx}(x,\sqrt{s})
&=& \int_{\gamma-i\infty}^{\gamma+i\infty}
\frac{dN}{2\pi i} x^{-N}
\sigma_H(N,\sqrt{s}).
\end{eqnarray}
As in a fixed-order calculation the Landau pole does not appear it is possible
to use any positive value of $\gamma$. Coincident results obtained at different
values of $\gamma$ prove the independence on its value.

\begin{figure}%
\centering
\parbox{2.9in}{%
\includegraphics[width=3in]{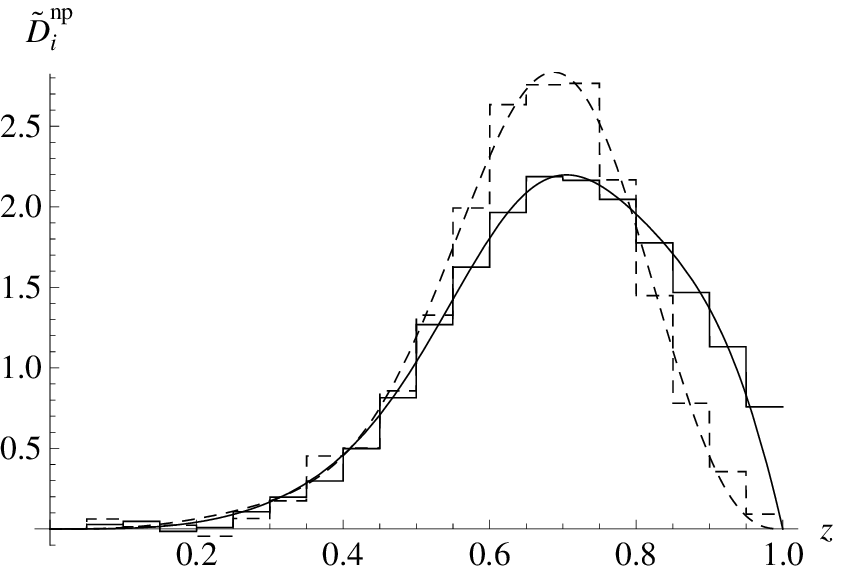}
\caption{\footnotesize
Nonperturbative fragmentation functions
of $D^*$-mesons and $\Lambda_C$-baryons together
with the approximative expressions (\ref{eq:npFF-klpM}) and (\ref{eq:npFF-klpB}).}%
\label{fig:numMBcmp}}%
\qquad
\begin{minipage}{2.9in}%
\includegraphics[width=3in]{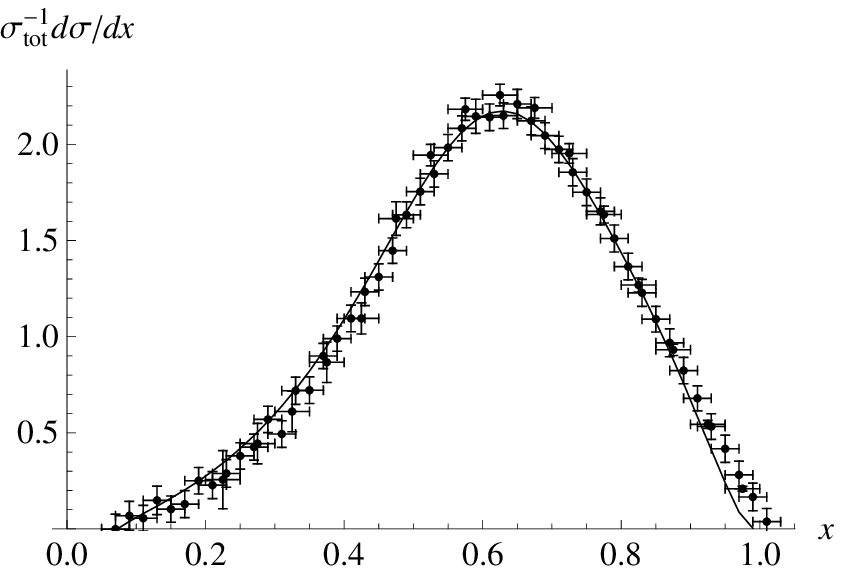}
\caption{\footnotesize $D^{*+}$-meson momentum distribution from the $c$-quark fragmentation
at $\sqrt{s}=10.58$~GeV energy together with the Belle and CLEO experimental data.}%
\label{fig:frag_DstP_Belle}%
\end{minipage}%
\end{figure}

\begin{figure}%
\centering
\parbox{2.9in}{%
\includegraphics[width=3in]{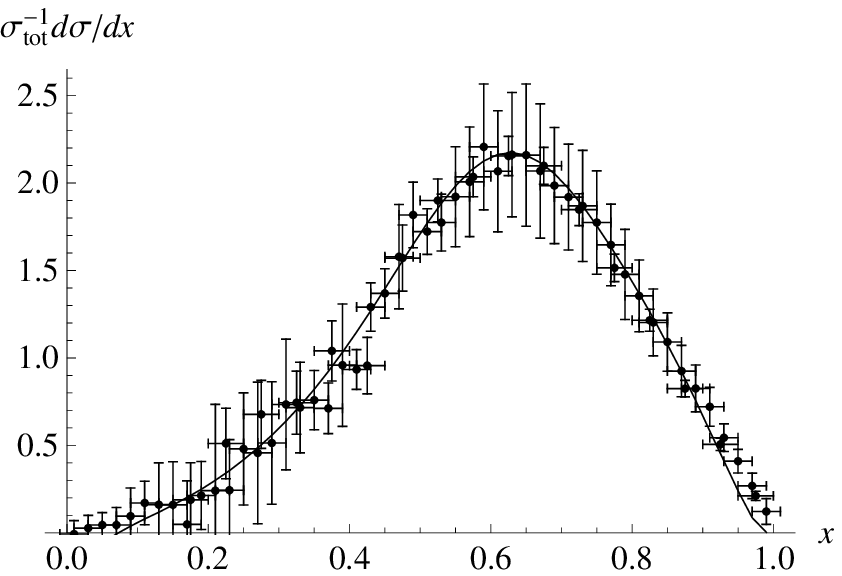}
\caption{\footnotesize $D^{*0}$-meson momentum distribution from the $c$-quark fragmentation
at $\sqrt{s}=10.58$~GeV energy together with the Belle and CLEO experimental data.}%
\label{fig:frag_Dst0_Belle}}%
\qquad
\begin{minipage}{2.9in}%
\includegraphics[width=3in]{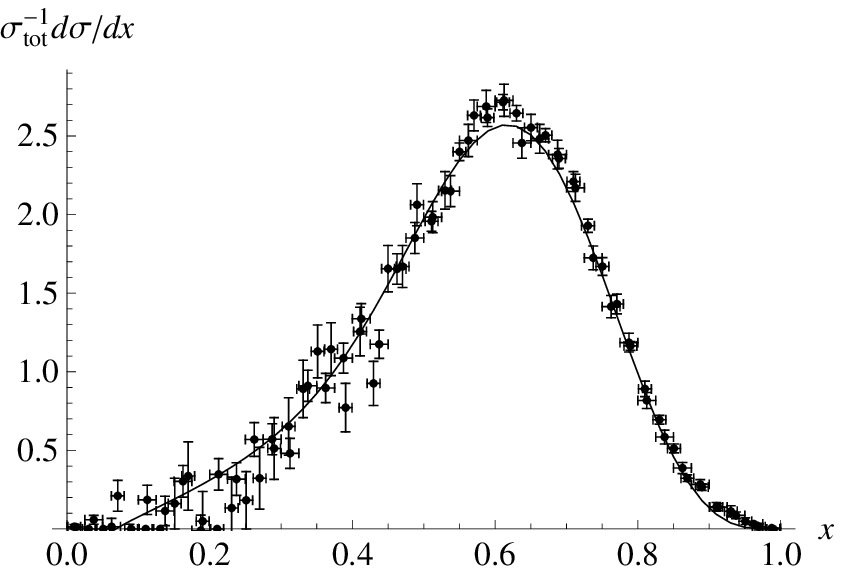}
\caption{\footnotesize $\Lambda_C$-baryon momentum distribution from the $c$-quark fragmentation
at $\sqrt{s}=10.58$~GeV energy together with the BaBar and Belle experimental data.}%
\label{fig:frag_Lc_Belle}%
\end{minipage}%
\end{figure}

Non-perturbative functions (\ref{eq:npFF-klpM}) and (\ref{eq:npFF-klpB}) allow to reproduce the
experimental data from $B$-factories with good precision, see
Fig. \ref{fig:frag_DstP_Belle}, \ref{fig:frag_Dst0_Belle} and  \ref{fig:frag_Lc_Belle}.

\begin{figure}%
\centering
\parbox{2.9in}{%
\includegraphics[width=3in]{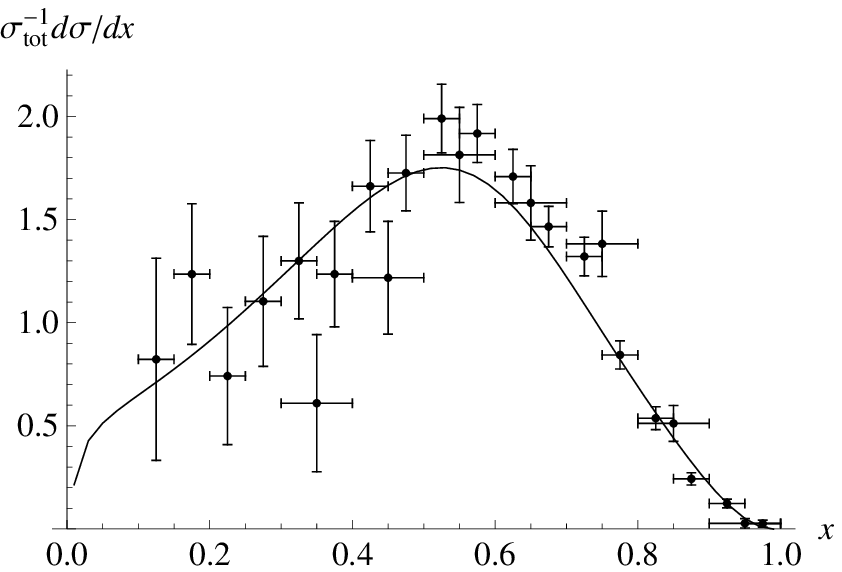}
\caption{\footnotesize $D^{*+}$-meson momentum distribution from the $c$-quark fragmentation
at $\sqrt{s}=91.18$~GeV energy together with the ALEPH and OPAL experimental data.}%
\label{fig:frag_D_LEP}}%
\qquad
\begin{minipage}{2.9in}%
\includegraphics[width=3in]{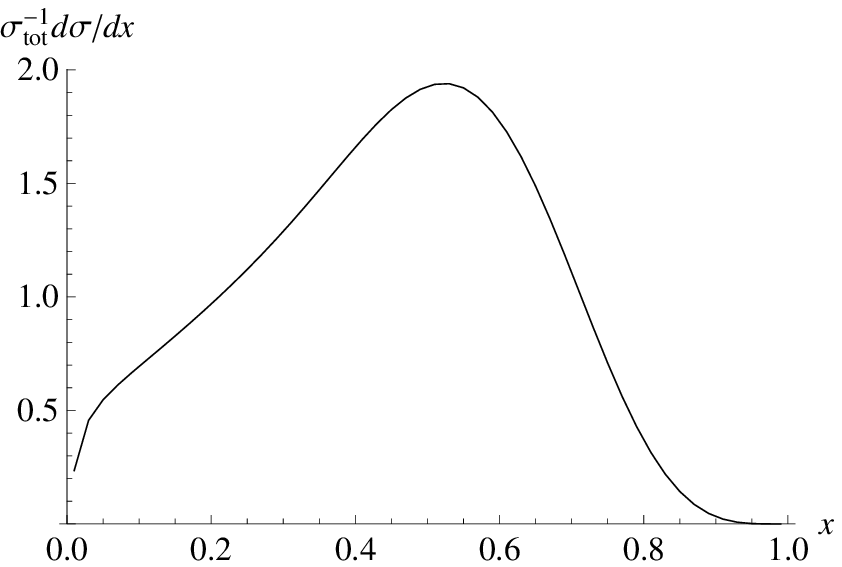}
\caption{\footnotesize Predicted $\Lambda_c$-baryon momentum distribution from the $c$-quark fragmentation
at $\sqrt{s}=91.18$~GeV energy.}%
\label{fig:frag_Lc_LEP}%
\end{minipage}%
\end{figure}

Evolution to the scale $\sqrt{s}/2=45.6$~GeV and non-perturbative
expression (\ref{eq:npFF-klpM}) are used to obtain the momentum distribution
of $D^*$-mesons at the $Z$-boson peak.
The predicted
spectrum together with the ALEPH and OPAL data is presented in Fig. \ref{fig:frag_D_LEP}.
These distributions coincide with the reasonable precision.
As the only difference in calculations for the $10$ and $90$ GeV energies
was in the final evolution scale, one can state that factorization relation
is valid in this energy range.

For the $\Lambda_c$ production at $91.2$~GeV energy the same evolution
in the perturbative component
is used. The non-perturbative
effects are described by the expression (\ref{eq:npFF-klpB}).
Unfortunately there is no experimental data on
the $\Lambda_c$ production at $Z$-boson peak. Our prediction
for the $\Lambda_c$ momentum distribution is presented in Fig. \ref{fig:frag_Lc_LEP}.

A considerable part of $D$'s is produced indirectly through
$D^*$ decays. The simple spin-state counting estimation $(2 J + 1)$
leads to the factor $3$ enchantment for $D^*$ production.
The experimental value obtained at $Z$-boson peak amounts to only $1.4$ \cite{Aubert:2006mp}.
We will use experimental value and assume that $D^{*+}$, $D^{*-}$, $D^{*0}$ and $\bar D^{*0}$
\footnote{
When further speaking about $D^{+}$ ($D^0$)
the same about $D^{-}$ ($\bar{D}^0$) should be implied.}
are produced with equal probabilities.

Following
the approach of \cite{Cacciari:2003zu} we assume that the $D$ meson
non-perturbative fragmentation function is the sum of a direct component, which
is isospin invariant plus the component arising from the $D^*$ decay.

The decay $D^*\to D\pi$ takes place very close to the threshold.
The momentum of $D$-meson in the $D^*$-meson rest system
\begin{eqnarray}
p' &=& \sqrt{\left({m_{D^*}^2 + m_D^2 - m_\pi^2 \over 2 m_{D^*}}\right)^2 - m_D^2} = 16~Β'
\end{eqnarray}
is sufficiently small to be neglected. Thus $D$ has the same
velocity as the $D^*$, and their momenta are thus proportional to their masses.
So the component of the $D$-meson fragmentation function
arising from $D^* \to D\pi$ decay is given by
\begin{eqnarray}
\tilde{D}^{D \pi}(z) &=& D^{D^*}_c\left(z {m_{D^*} \over {m_{D}}}\right) \,
\theta\left(1 - z {m_{D^*} \over {m_{D}}}\right){m_{D^*} \over {m_{D}}},
\label{eq:Dst2Dpi}
\end{eqnarray}
where $D^{D^*}_c(z)$ is the non-perturbative fragmentation function of
$D^*$-meson (\ref{eq:npFF-klpM}). The integral of the expression (\ref{eq:Dst2Dpi})
equals 1, so if should enter the $D$-meson fragmentation function with the
weight proportional to the $D^* \to D \pi$ decay probability and probability
of the $D^*$ production.

What concerns the $D^*\to D\gamma$ decay,
momentum of the $D$ in the $D^*$ frame is non-negligible:
\begin{eqnarray}
p' &=& {m_{D^*}^2 - m_D^2 \over 2 m_{D^*}} = 135~Β'.
\end{eqnarray}
The
$D$ momentum in the laboratory frame is given by
a Lorentz boost
\begin{eqnarray}
p &=& \gamma ( p'\cos\theta+\beta \epsilon' ),
\end{eqnarray}
where $\beta$ is the velocity of the $D^*$-meson, $\gamma = 1 / \sqrt{1 - \beta^2}$,
$\epsilon' = {(m_{D^*}^2 + m_D^2) / (2 m_{D^*})}$ --- energy of the $D$-meson
in the $D^*$ rest frame and $\theta$ --- its decay angle with respect to the $D^*$ direction.
Denoting momentum and energy of the $D^*$ in laboratory frame by $p^*$ and $\epsilon^*$
respectively one obtains
\begin{eqnarray}
\gamma &=& { \epsilon^* \over {m_{D^*}} }, \;
\beta = { p^* \over {\gamma m_{D^*}} }.
\end{eqnarray}
Introducing variables
\begin{eqnarray}
z &=& { p \over p^D_{\rm max} } \equiv { p \over \sqrt{s/4 - m_D^2} } ,
 \nonumber \\
z^* &=& { p^* \over p^{D^*}_{\rm max} } \equiv { p^* \over \sqrt{s/4 - m_{D^*}^2} },
\end{eqnarray}
contribution of the $D^* \to D \gamma$ decay to the $D$ production
can be written down as
\begin{eqnarray}
\tilde{D}^{D \gamma}(z) &=& \int_0^1 {d {z^*}} \int_{-1}^1 { {d \cos \theta} \over 2} \,
D^{D^*}_c( z^* ) \, \delta \left( z - \gamma \, { p'\cos\theta+\beta \epsilon'  \over {p_{\rm max}}} \right).
\label{eq:Dst2Dph}
\end{eqnarray}
As in the previous case the integral of this expression is normalized to unity.

The branching ratios involved are \cite{Amsler:2008zzb}:
\begin{eqnarray}
  Br_{D^{*+} \to D^0 \pi^+} &=& 67.7 \pm 0.5, \% \nonumber \\
  Br_{D^{*+} \to D^+ \pi^0} &=& 30.7 \pm 0.5, \% \nonumber \\
  Br_{D^{*+} \to D^+ \gamma} &=& 1.6 \pm 0.4, \%  \\
  Br_{D^{*0} \to D^0 \pi^0} &=& 61.9 \pm 2.9, \% \nonumber \\
  Br_{D^{*0} \to D^0 \gamma} &=& 38.1 \pm 2.9. \% \nonumber
\end{eqnarray}

Finally non-perturbative fragmentation functions of $D$-mesons can be
written down as follows:
\begin{eqnarray}
\tilde{D}^{D^+}_c(z) &=& n^{D^+} (
D^{D}_c(z) + c \left[ Br_{D^{*+} \to D^+ \gamma} \tilde{D}^{D \gamma}(z) + \right. \nonumber \\
&+& \left. Br_{D^{*+} \to D^+ \pi^0} \tilde{D}^{D \pi}(z) \right] )
\label{eq:FF-D+}
\end{eqnarray}
and
\begin{eqnarray}
\tilde{D}^{D^0}_c(z) &=& n^{D^0} (
D^{D}_c(z) + c \left[ Br_{D^{*0} \to D^0 \gamma} \tilde{D}^{D \gamma}(z) +  \right. \nonumber \\
&+& \left. (Br_{D^{*+} \to D^0 \pi^+} + Br_{D^{*0} \to D^0 \pi^0}) \tilde{D}^{D \pi}(z) \right] ),
\label{eq:FF-D0}
\end{eqnarray}
where $c = 1.4$ --- ratio of probabilities to fragment into $D^*$ and $D$-mesons,
coefficients $n^{D^+}$ and $n^{D^0}$ provide normalization of fragmentation functions to unity:
\begin{eqnarray}
n^{D^+} &=& (1+c(Br_{D^{*+} \to D^+ \gamma} + Br_{D^{*+} \to D^+ \pi^0} ) )^{-1}, \nonumber \\
n^{D^0} &=& (1+c(Br_{D^{*0} \to D^0 \gamma} + Br_{D^{*+} \to D^0 \pi^+} + Br_{D^{*0} \to D^0 \pi^0}))^{-1}.
\end{eqnarray}

\begin{figure}%
\centering
\parbox{2.9in}{%
\includegraphics[width=3in]{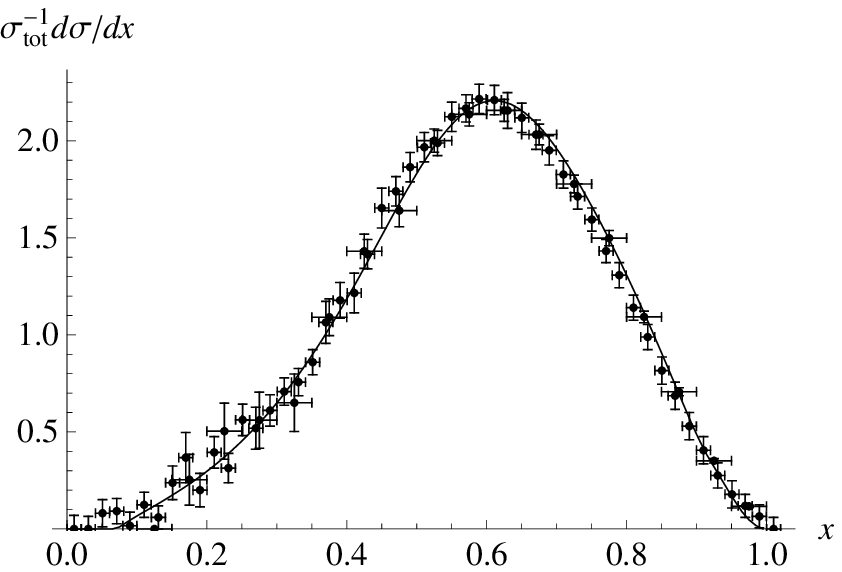}\\
\caption{\footnotesize $D^+$-meson momentum distribution from the $c$-quark fragmentation
at $\sqrt{s}=10.58$~GeV energy together with the Belle and CLEO experimental data.}
\label{fig:frag_D+_Belle}}
\qquad
\begin{minipage}{2.9in}%
\includegraphics[width=3in]{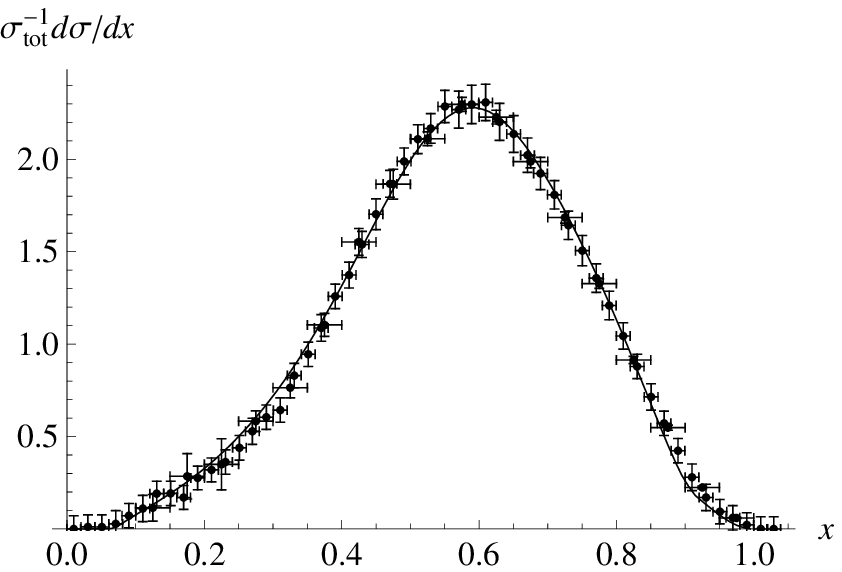}\\
\caption{\footnotesize $D^0$-meson momentum distribution from the $c$-quark fragmentation
at $\sqrt{s}=10.58$~GeV energy together with the Belle and CLEO experimental data.}
\label{fig:frag_D0_Belle}
\end{minipage}%
\end{figure}

In order to obtain momentum spectra of $D$-mesons expression (\ref{eq:sigm-x})
is used. The result for $D^+$-mesons is presented in Fig. \ref{fig:frag_D+_Belle},
for $D^0$-mesons --- in Fig. \ref{fig:frag_D0_Belle}.
Both distributions are in a good agreement with experimental data.
The $D^0$ spectrum is slightly softer then the $D^+$ one because of the
larger probability of the $D^* \to D^0 X$ decay.

\section{Charmed hadron production in $B$-meson decays}

Let us now consider charmed particles production in the $b$-quark decays.
The energy of $10.58~\rm{GeV}$ corresponds to the $\Upsilon(4S)$ resonance
which decays into a $B \bar B$-pair with almost unitary probability.
$B$-mesons from such decays are nearly at rest as their mass
$m_B = 5.28~\rm{GeV} \simeq \sqrt{s}/2$. Neglecting $b$-quark motion
within the meson the $c$-quark spectrum from $B$ decay can be
easily found. The charmed hadron spectrum can then be written down as
\begin{eqnarray}
{{d\sigma_H}\over{dx}} (x) &=& \int_{x}^1 {{{dz}\over z} \left({{d\sigma_{b \to c}}\over {dz}}(z)\right)
{D^{\rm np}_c}\left({{x} \over {z}}\right)},
\label{eq:H-spect-B}
\end{eqnarray}
where ${d\sigma_{b \to c}} / {dz}$ is the $c$-quark spectrum
obtained by analysis of weak decays
$b \to c + l \nu_l$, $b \to c + q \bar q$,
$b \to c + \bar c s$ and $\bar b \to \bar c + c \bar s$.

We neglect here any perturbative fragmentation function
as at such low energy it should not sufficiently differ
from the $\delta$-function. The non-perturbative component
of $D^*$ production
is described by the expression (\ref{eq:npFF-klpM}).

In experimental data value $x = 1$ corresponds to the largest
possible momentum of the $D^*$-meson prodused at $10.58~\rm{GeV}$ energy:
\begin{eqnarray}
p^{D^*}_{\rm max}=\sqrt{s/4-m_{D^*}^2}=4.88~\rm{GeV}
\label{eq:pHmax-D}
\end{eqnarray}
while the largest possible momentum of the $D^*$-meson from the
$B$ decay is equal to the largest $c$-quark momentum
\begin{eqnarray}
p^{c}_{\rm max}=\sqrt{{{(m_b^2+m_c^2)^2}\over{4 m_b^2}} - m_c^2}=2.24~\rm{GeV}.
\label{eq:pcmax-D}
\end{eqnarray}
This value actually coincides with those calculated via the hadron (not parton) masses:
\begin{eqnarray}
p^{B \to D^*}_{\rm max}=\sqrt{{{(m_B^2+m_D^2)^2}\over{4 m_B^2}} - m_D^2}=2.26~\rm{GeV}.
\label{eq:pDmax-D}
\end{eqnarray}

\begin{figure}%
\centering
\parbox{2.9in}{%
\includegraphics[width=3in]{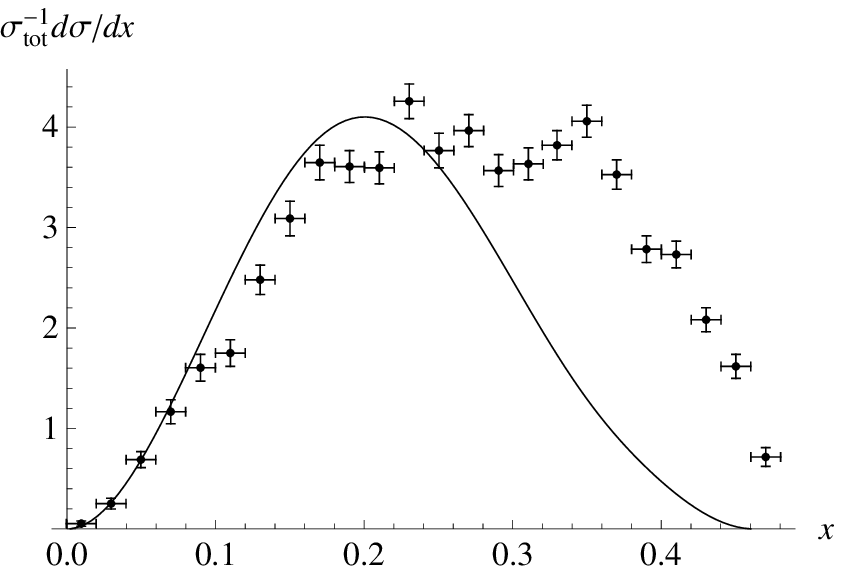}
\caption{\footnotesize $D^*$-meson momentum distribution from the $b$-quark fragmentation
at $\sqrt{s}=10.58$~GeV energy together with the Belle experimental data.}%
\label{fig:frag_D_Belle_U}}%
\qquad
\begin{minipage}{2.9in}%
\includegraphics[width=3in]{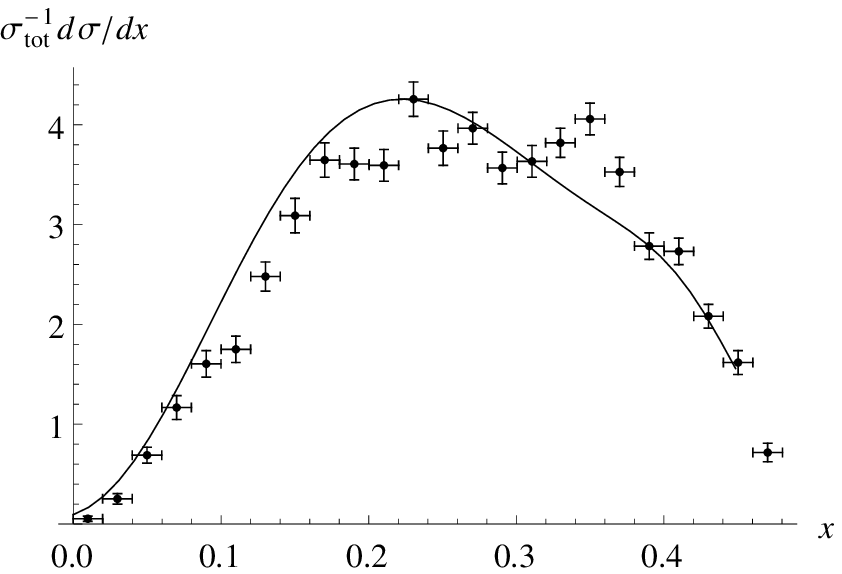}
\caption{\footnotesize $D^*$-meson momentum distribution from the $b$-quark fragmentation and
recombination with the light quark from $B$-meson
at $\sqrt{s}=10.58$~GeV energy together with the Belle experimental data.}%
\label{fig:frag_D_Belle_U_h}%
\end{minipage}%
\end{figure}

Thus the momentum distribution of $D$-mesons from the $B$ decays is located in the region
$x < p^{c}_{\rm max}/p^{D^*}_{\rm max} = 0.46$. The distribution concerned and the
experimental data are plotted in Fig. \ref{fig:frag_D_Belle_U}. In this only plot the normalization of the
experimental data was changed for points in the low $x$ region to be in agreement with the predictions.
In the high $x$ region there is some extra contribution not predicted by the
fragmentation model. We suppose that this contribution arises from the interaction
of the charm quark with the light valent quark from the $B$-meson. As this quark has
larger momentum then a sea quark the process concerned contributes to the
high $x$ region. We assume that the momentum distribution of $D^{*}$-mesons
produced in this process coincides with the $c$-quark distribution.
This assumption allows to obtain better agreement with the experimental data (Fig. \ref{fig:frag_D_Belle_U_h}).

Let us now proceed to the $\Lambda_C$-baryon production. To obtain
$\Lambda_C$-baryon momentum distribution one should convolute $c$-quark
spectrum with the non-perturbative fragmentation function
(\ref{eq:npFF-klpB}). The largest
possible momentum of the $\Lambda_C$-baryon prodused
at $10.58~\rm{GeV}$ energy is equal
\begin{eqnarray}
p^{\Lambda_c}_{\rm max}=\sqrt{s/4-m_{\Lambda_c}^2}=4.76~Ē'.
\label{eq:pHmax-L}
\end{eqnarray}

When finding largest $\Lambda_C$-baryon momentum from the $B$ decay
it is important to mention that baryon production is always
accompanied by the any-baryon emergence. The lightest of them is
anti-proton. Thus the largest possible $\Lambda_C$-baryon momentum
equals to
\begin{eqnarray}
p^{B \to \Lambda_c}_{\rm max}=\sqrt{{{(m_B^2+m_{\Lambda_c}^2-m_p^2)^2}\over{4 m_B^2}} - m_{\Lambda_c}^2}=2.02~Ē'.
\label{eq:pDmax-D}
\end{eqnarray}
This value differs significantly from the $p^{c}_{\rm max}$.

\begin{figure}
  \includegraphics[width=3in]{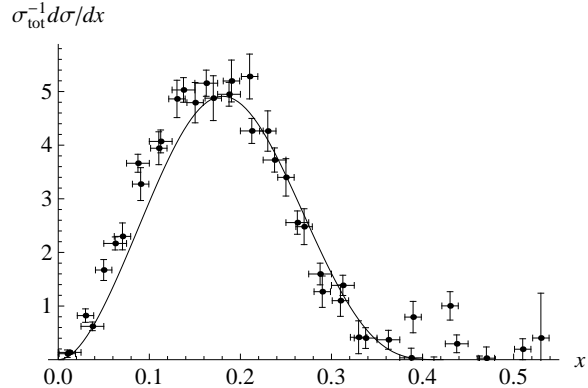}\\
  \caption{\footnotesize $\Lambda_C$-baryon momentum distribution from the $b$-quark fragmentation
at $\sqrt{s}=10.58$~GeV energy together with the BaBar and Belle experimental data.}
  \label{fig:frag_Lc_Belle_U}
\end{figure}

Thereby the distribution sought for occupies the region
$x < p^{B \to \Lambda_c}_{\rm max}/p^{\Lambda_c}_{\rm max} = 0.42$.
It reasonably coincides with the experimental data (Fig. \ref{fig:frag_Lc_Belle_U}).
As opposite to the meson production
sea $ud$-diquark capturing in needed for the $\Lambda_C$ formation is needed.
Tat is why valent quark from the $B$-meson does not change the prediction
of fragmentation approach in case of baryon production.

\section{Conclusion}

In this article charmed hadron production in the vast range of
energies was concerned. During this the energy dependence was contained only
in the perturbative component of fragmentation function, for
which the NLO-expression was used. Good agreement with the experimental
data points to the fulfilment of the factorization assumption in the $10$ to $90$ GeV
energy range.

It is important to mention that non-perturbative fragmentation functions
were one and the same for all energies except $D^*$-mesons production
in $B$ decays.
It means that separation of non-perturbative phenomena was carried out correctly.
Indeed the perturbative part is relevant for charmed quark production only,
the non-perturbative --- only for its transaction to the final particle.

The difference in meson and baryon production in the $x \to 1$ region
dues to the different non-perturbative fragmentation functions. The difference
in them by-turn is explained by quark-counting. This phenomenon takes
place in large $x$ region only, where non-perturbative phenomena is most
significant.

In the $x \to 0$ region non-perturbative fragmentation functions of charmed
mesons and baryons coincide. This is relevant to the one and the same behavior
of charm quark distribution functions in mesons and baryons in the low-$x$ limit.
Moreover, the parameter $\alpha$ which determines the low-$x$ behavior is close
to the estimations of the $\alpha_c$ --- interception of the charmed Regge trajectory.

At the low energy scale which corresponds to the $B$-meson mass a discrepancy in the
$D^*$ momentum distribution was found. The origin of it is thought to due mainly
to the influence of a valent quark from decaying $B$-meson. Such discrepancy does
not occur in $\Lambda_C$ production which is in a pretty good agreement with the
experimental data. The $m/\sqrt{s}$-corrections to the factorization relation
also worth mentioning at the $m_B$ energy scale.

Author would like to thank Prof. Likhoded A.K. and Luchinsky A.V. for useful ideas and discussions.
The work was financially supported by Russian Foundation for Basic Research (grant\#10-02-00061-a)
and grants of the president of Russian Federation MK-140.2009.2 and MK-406.2010.2.

\end{document}